\documentclass{IEEEtran}
\ifCLASSINFOpdf
\else
\fi
\usepackage[process=auto,cleanup={.dvi,.ps,.pdf,.log}]{pstool}
\begingroup
\makeatletter
\catcode`\#=11
\gdef\pstool@bitmap@opts{%
	-dAutoFilterColorImages#false
	-dAutoFilterGrayImages#false %
	-dColorImageFilter#/FlateEncode %
	-dGrayImageFilter#/FlateEncode 
}
\gdef\pstool@pspdf@opts{%
	-dPDFSETTINGS#/prepress %
	-dCompatibilityLevel#1.3 %
	-dEmbedAllFonts#true %
	-dSubsetFonts#true
}
\endgroup
\graphicspath{{figures/}}

\usepackage[caption=false, font=footnotesize,width=8cm,nearskip=5pt]{subfig}

\usepackage{amsfonts}
\PassOptionsToPackage{hyphens}{url}

\renewcommand{\vec}[1]{\ensuremath{\boldsymbol{#1}}} 
\usepackage[cmex10]{amsmath}
\usepackage{booktabs}
\newcommand{\PreserveBackslash}[1]{\let\temp=\\#1\let\\=\temp}
\let\PBS=\PreserveBackslash
\usepackage{multicol}
\usepackage{multirow}
\usepackage{longtable}
\usepackage{tabularx}
\usepackage{tabulary}
\usepackage{amsfonts}
\usepackage{blkarray, bigstrut} %

\begin{document}
%
\title{Transmission Network Reduction Method using Nonlinear Optimization}



 \author{\IEEEauthorblockN{Philipp Fortenbacher\IEEEauthorrefmark{1},
 Turhan Demiray\IEEEauthorrefmark{1},
 Christian Schaffner\IEEEauthorrefmark{2}}
 \IEEEauthorblockA{\IEEEauthorrefmark{1} Research Center for Energy Networks (FEN)\\
 ETH Zurich, Switzerland \\ \{fortenbacher, demirayt\}@fen.ethz.ch}
 \IEEEauthorblockA{\IEEEauthorrefmark{2} Energy Science Center (ESC)\\
 ETH Zurich, Switzerland \\ schaffner@esc.ethz.ch}
 }

\maketitle

\begin{abstract}
This paper presents a new method to determine the susceptances of a reduced transmission network representation by using nonlinear optimization. We use Power Transfer Distribution Factors (PTDFs) to convert the original grid into a reduced version, from  which we determine the susceptances. From our case studies we find that considering a reduced injection-independent evaluated PTDF matrix is the best approximation and is by far better than an injection-dependent evaluated PTDF matrix over a given set of arbitrarily-chosen power injection scenarios. We also compare our nonlinear approach with existing methods from literature in terms of the approximation error and computation time. On average, we find that our approach reduces the mean error of the power flow deviations between the original power system and its reduced version, while achieving higher but reasonable computation times.    
\end{abstract}


\section{Introduction}
Network reduction has become important for power system planning and operation, since a reduced version of a large network is required to cope with the computational burden in optimal planning and unit commitment (UC) applications. Although some methods make use of decomposition methods, optimal network expansion planning \cite{DelaTorre2008}, joint optimal generation and network expansion planning \cite{Baringo2012,Roh2007,AlvarezLopez2007} studies or UC problems \cite{Murillo-Sannchez2013} usually incorporate fairly small networks. Even decomposing the problem structure and running the problems on high performance computing clusters require a reduced version of the original network \cite{Papavasiliou2013}. This is due to the fact that such optimization problems are inter-temporal coupled in the form of multi-period optimal power flow (OPF) problems. Their computational complexity depends at least polynomially on the size of the network, the number of generators, and the time horizon \cite{isgt,Fortenbacher2017}. Moreover, network reduction methods could also be applied in flow-based power market clearing schemes \cite{Bergh2015} to reduce the amount of critical branches in the original network topology or to find a concentrated network representation with respect to the bidding zones. 

The first corner stone in network reduction was lain by Ward \cite{Ward1949} with the introduction of circuit equivalents. Following-up works \cite{Tinney1987,Shi2012} still preserve the circuit equivalents with more enhanced adaptive equivalencing techniques. These approaches consider a partitioning of internal, external, and boundary bus groups. While the internal system remains unchanged, the external and boundary buses are transformed to represent the electrical system. However, this might lead to ill-suited systems when one wants to retain the system e.g. with respect to geographical representation.

The works \cite{Cheng2005,Oh2010,Shi2015} improve this drawback by projecting the reduced Power Transfer Distribution Factor (PTDF) matrix on the full original PTDF under a zonal bus assignment. It is noteworthy, that this conversion is a lossy approximation. While the reduced PTDF matrix of \cite{Shi2015} depends on the operational setpoint, i.e. the PTDF matrix is power injection-dependent, the author in \cite{Oh2010} derived a reduced PTDF matrix which is injection-independent. Since all presented applications vary generation and load setpoints, they require to be independent from the power injection point. 

Reduced power system representations relying only on PTDFs induce still a high computational complexity to multi-period OPF problems, since the PTDF matrix is dense. To allow for tractable multi-period problems it is crucial to have sparse network models. This property is fulfilled by the nodal admittance matrix containing the susceptances of the transmission network. The authors of~\cite{Oh2010,Shi2015} determine these quantities based on the reduced PTDF via an Eigenvalue decomposition or a linear least-squares fitting problem introducing additional power flow errors between the original and reduced power flow solution.
  
In sum, this means that there is a clear need for sparse and injection-independent equivalents that introduce low power flow errors between the original and the reduced system. Hence, the objective of this paper is to develop a new method that reduces this error and to compare this approach with existing methods from literature.

The contribution of this paper is the development of a new method that is based on nonlinear optimization to determine the susceptances of the reduced PTDF matrix. We compare our method with existing methods over a range of different power injection scenarios and differently-sized grids in terms of the power flow error and computation time.

The remainder of this paper is organized as follows. Section~\ref{sec:method} reviews existing PTDF-based network reduction methods and describes our new susceptance determination method. Section~\ref{sec:results} presents two case studies comparing the power flow approximation errors and computation times illustratively for the IEEE14 grid and differently-sized grids. Section~\ref{sec:conclusion} draws the conclusion.

\section{Method}
\label{sec:method}
\begin{figure}[t]
	\centering
	\def\svgwidth{\columnwidth}
	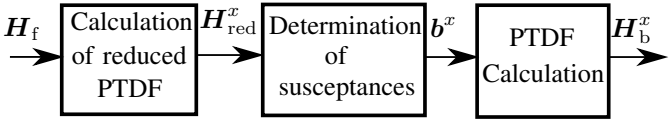
	\caption{Process of finding the a sparse reduced network formulation. The first two stages are needed to determine the susceptances, while the third one is necessary for the validation.}
	\label{fig:process}
\end{figure}

In this section we first review and compare existing PTDF-based network reduction techniques and derive the reduced PTDF matrices in a common framework. Second, we present our new method to determine the susceptances. In general, finding a sparse reduced network formulation comprises a two stage process and a following validation process (see Fig.~\ref{fig:process}). The first stage calculates the reduced PTDF $\vec{H}_\mathrm{red}^x$ based on the full PTDF $\vec{H}_\mathrm{f}$ from the original system. Considering a full network with $n_b$ buses and $n_l$ branches, the full PTDF matrix $\vec{H}_\mathrm{f}\in \mathbb{R}^{n_l \times n_b-1}$ maps the bus injections $ \vec{p}_\mathrm{inj} \in \mathbb{R}^{n_b-1 \times 1}$ to the line flows $\vec{p}_\mathrm{f} \in \mathbb{R}^{n_l \times 1}$ as follows
\begin{equation}
\vec{p}_\mathrm{f} = \underbrace{\vec{B}_\mathrm{br}{\vec{B}_\mathrm{bus}}^{-1}}_{:= \vec{H}_\mathrm{f}}\vec{p}_\mathrm{inj} \quad, \label{eq:Hf1}
\end{equation}
\noindent where $\vec{B}_\mathrm{br} \in \mathbb{R}^{n_l\times n_b-1}$ is the slack bus adjusted branch susceptance matrix and $\vec{B}_\mathrm{bus} \in \mathbb{R}^{n_b-1\times n_b-1}$ is the slack bus adjusted nodal susceptance matrix. The entries corresponding to the slack bus are deleted, in order to avoid that the matrix inverse gets singular, since the full version of $\vec{B}_\mathrm{bus} \in \mathbb{R}^{n_b\times n_b}$ is rank-deficient. 

The second stage determines the virtual susceptances $\vec{b}^x \in \mathbb{R}^{n_l^r \times 1}$ from the reduced PTDF or using the physical information of the grid structure. The variable $n_l^r$ specifies the number of the zonal interconnections.

The third stage is needed to validate the network reduction results. We use here the fact that the reduced PTDF can be alternatively determined by using the susceptance information. E.g. the PTDF matrix of the full system is given by
\begin{equation}
\vec{H}_\mathrm{f} = \mathrm{diag}\{\vec{b}\}\vec{C}_\mathrm{f} \left(\vec{C}_\mathrm{f}^T\mathrm{diag}\{\vec{b}\}\vec{C}_\mathrm{f} \right)^{-1} \quad, \label{eq:Hf2}
\end{equation}
\noindent where $\vec{C}_\mathrm{f} \in \mathbb{Z}^{n_l \times n_b-1} $ is the node-branch incidence matrix and $\vec{b} \in \mathbb{R}^{n_l \times 1} $ are the susceptances of the original system. 

We further detail the corresponding stages in the following subsections.

\subsection{Calculation of reduced PTDF}
To find the susceptances of the reduced network, we first need to calculate the reduced PTDF of the original network. Here, we compare and review two versions from literature \cite{Oh2010,Shi2015} that are expressed and derived in our notation framework. The reduced network is defined by a set of zones. Each zone specifies a set of original buses belonging to the individual zone. This can be achieved by constructing the bus to zone incidence matrix $\vec{T}_\mathrm{bz} \in \mathbb{Z}^{n_z-1 \times n_b-1}$, where $n_z$ defines the number of zones. The interzonal powerflows $\vec{p}_\mathrm{f}^\mathrm{red} \in \mathbb{R}^{n_l^r \times 1}$  can be expressed as 
\begin{equation}
\vec{p}_\mathrm{f}^\mathrm{red} = \vec{H}_\mathrm{red}\vec{T}_\mathrm{bz} \vec{p}_\mathrm{inj} \quad, \label{eq:p_red2}
\end{equation}
\noindent where the reduced PTDF matrix $\vec{H}_\mathrm{red} \in \mathbb{R}^{n_l^r \times n_z-1}$ is also slack-bus adjusted.

Under the zonal assignment $\vec{T}_\mathrm{bz}$ we can also calculate the inter-zonal power flows by
\begin{equation}
\vec{p}_\mathrm{f}^\mathrm{red} = \vec{T}_\mathrm{f}\vec{H}_\mathrm{f} \vec{p}_\mathrm{inj} \quad, \label{eq:p_red}
\end{equation}  
\noindent where $\vec{T}_\mathrm{f} \in \mathbb{Z}^{n_l^r \times n_l} $ is the original branch to zonal interconnection matrix, which sums up the original corresponding power flows to the zonal power flows.

\subsubsection{Injection-independent PTDF}
By equating \eqref{eq:p_red} and \eqref{eq:p_red2}, canceling $\vec{p}_\mathrm{inj}$, and minimizing for $\vec{H}_\mathrm{red}$, it follows that
\begin{equation}
\vec{H}_\mathrm{red}^{\mathrm{ind}} = \vec{T}_\mathrm{f}\vec{H}_\mathrm{f}\vec{T}_\mathrm{bz}^T\left(\vec{T}_\mathrm{bz}\vec{T}_\mathrm{bz}^T \right)^{-1} \quad. \label{eq:redPTDFoh}
\end{equation}
\noindent This is the result of \cite{Oh2010}, but written in a more compact form. Note that $\vec{H}_\mathrm{red}^{\mathrm{ind}}$ is injection-independent due to the cancelation of $\vec{p}_\mathrm{inj}$.     

\subsubsection{Injection-dependent PTDF}
If we diagonalize $\vec{p}_\mathrm{inj}$, multiply \eqref{eq:p_red} and \eqref{eq:p_red2} with $\vec{T}_\mathrm{bz}^T$, and equating \eqref{eq:p_red} and \eqref{eq:p_red2}, we can write 
\begin{equation}
 \vec{H}_\mathrm{red}\vec{T}_\mathrm{bz} \text{diag}  \{\vec{p}_\mathrm{inj} \} \vec{T}_\mathrm{bz}^T = \vec{T}_\mathrm{f}\vec{H}_\mathrm{f} \text{diag}  \{\vec{p}_\mathrm{inj} \} \vec{T}_\mathrm{bz}^T \quad.
\end{equation}

\noindent Obtaining the injection-dependent reduced PTDF $\vec{H}_\mathrm{red}^{\mathrm{dep}}$ in a Least Squares sense, can be done by applying
\begin{equation}
\vec{H}_\mathrm{red}^{\mathrm{dep}} = \vec{T}_\mathrm{f}\vec{H}_\mathrm{f} \text{diag}  \{\vec{p}_\mathrm{inj} \}\vec{T}_\mathrm{bz}^T  \left(\text{diag}  \{\vec{T}_\mathrm{bz}\vec{p}_\mathrm{inj} \}\right)^{-1}\quad, \label{eq:redPTDFshi}
\end{equation}

\noindent which is the identical result from \cite{Shi2015}.

\subsection{Determination of Susceptances}
\label{sec:detSus}
Once the reduced PTDF is calculated, we can use this result to determine the susceptances. Four methods will be discussed in this paper:  

\subsubsection{Physical B Approximation}
This method uses only the physical information of the grid structure. By applying 
\begin{equation}
\vec{b}^\mathrm{phys} = \vec{C}_\mathrm{phys} \vec{b} \quad,
\end{equation}
\noindent we can map the physical original susceptances $\vec{b}$ to the zonal interconnections $\vec{b}^\mathrm{phys}$. The matrix  $\vec{C}_\mathrm{phys} \in \mathbb{R}^{n_l \times n_l^r} $ sums up the corresponding original branch susceptances that belong to the interconnected zones. It is noteworthy that this method cannot capture any intrazonal impacts on the interzonal power flows. This aspect translates to higher approximation errors on interzonal power flows.

\subsubsection{Least Squares Fitting}
\label{sec:lsfit}
For comparison reasons, we review the method from \cite{Shi2015}. It is based on a Least Squares (LS) fitting problem that minimizes the Euclidean distance of 
\begin{equation}
\min_{\vec{b}^\mathrm{shi}}\left\| \left[ \begin{array}{c}
\vec{a}^T \\
\vec{\Theta}
\end{array} \right] \vec{b}^\mathrm{shi} - \left[ \begin{array}{c}
b_m \\
\vec{0}
\end{array} \right] \right\|_2 \quad,
\label{eq:lsmin}
\end{equation}

\noindent where
\begin{equation}
\vec{\Theta} = \left[ \begin{array}{c}
(\vec{H}_\mathrm{red} \vec{C}^T - \vec{I}) \text{diag}\{\vec{c}_1\} \\
\vdots \\
(\vec{H}_\mathrm{red} \vec{C}^T - \vec{I}) \text{diag}\{\vec{c}_{n_z-1}\} 
\end{array} \right] \quad .
\end{equation}

The vector $\vec{b}^\mathrm{shi}$ specifies the virtual susceptances of the interzonal connections, the matrix $\vec{C} \in \mathbb{Z}^{n_l^r \times n_z -1} = [\vec{c}_1,\ldots,\vec{c}_{n_z-1}]$ is the node-branch incidence matrix of the reduced network and $\vec{I}$ is the Identity. Since any constant applied to $\vec{b}^\mathrm{shi}$ solves the equation $\vec{\Theta}\vec{b}^\mathrm{shi} = \vec{0}$ and therefore induces multiple solutions, we fix the solution that has a physical meaning. In this way, we select the strongest physical interconnection by  $b_m = \max(\vec{b}^\mathrm{phys})$ and introduce the column vector $\vec{a}^T\in \mathbb{Z}^{1 \times n_l^r } $, in which $a_m =1$ and the remaining elements are set to zero.

\subsubsection{Eigenvalue Decomposition}
\label{sec:evd}
As described in \cite{Oh2010} it is also possible to determine the susceptances by using an Eigenvalue decomposition. We can define the following decomposition  
\begin{equation}
\vec{H}_\mathrm{red}\vec{C}^T = \vec{V}\vec{\Lambda}\vec{V}^{-1} \quad,
\end{equation}
\noindent where $\vec{V}$ are the Eigenvectors with respect to the diagonal Eigenvalue matrix $\vec{\Lambda}$. The unity eigenvectors $\vec{E}$ of $\vec{V}$ are appended to $\vec{V}$. After performing following QR decomposition
\begin{equation}
[\vec{V} \ \vec{E} ] = [\vec{Q}_1 \ \vec{Q}_2] \left[\begin{array}{c} \vec{R}_1 \\ \vec{0} \end{array}\right] \quad,
\end{equation}
\noindent the matrix $\vec{Q}_2$ is used to construct the following matrix $\vec{\Omega}$
\begin{equation}
\vec{\Omega} = \left[ \begin{array}{c}
\vec{Q}_2 \text{diag}\{\vec{c}_1\} \\
\vdots \\
\vec{Q}_2 \text{diag}\{\vec{c}_{n_z-1}\}
\end{array} \right] \quad.
\end{equation}
\noindent With the following LS estimation
\begin{equation}
\begin{array}{l}
\displaystyle\min_{\vec{b}^\mathrm{oh}} \ \left \| \vec{\Omega}\vec{b}^\mathrm{oh} \right\|_2\\
\text{s.t.} \\
\| \vec{b}^\mathrm{oh} \|_2 \geq M \quad,
\end{array}
\end{equation}
\noindent we obtain the interzonal susceptances $\vec{b}^\mathrm{oh}$. Apart from the Eigenvalue decomposition the difference between the aforementioned method is that the 2-norm of  $\vec{b}^\mathrm{oh}$ is set to a small positive number $M$ as a lower bound in the optimization problem.   

\subsubsection{Nonlinear Optimization}
The methods in Sections~\ref{sec:lsfit}~--~\ref{sec:evd} rearrange the B approximation problem into a linear system of equations. The corresponding overdetermined linear systems are then solved for the susceptances by LS fitting in a vector norm. Instead of using a vector norm, the main idea here is to formulate an optimization problem in a more direct approach minimizing a matrix norm metric to achieve a better performance. Consider that \eqref{eq:Hf2} does also hold for the reduced network:
\begin{equation}
\vec{H}_\mathrm{red} = \mathrm{diag}\{\vec{b}\}\vec{C} \left(\vec{C}^T\mathrm{diag}\{\vec{b}\}\vec{C} \right)^{-1} \quad, \label{eq:Hred2}
\end{equation}
\noindent where $\vec{C}$ is the node-branch incidence matrix and captures the topology of the reduced network. For the optimization it is fixed and is defined a-priori by the zonal assignment. We already have an expression for $\vec{H}_\mathrm{red}$ by applying \eqref{eq:redPTDFoh} or \eqref{eq:redPTDFshi}, such that we only need to find an optimal susceptance vector $\vec{b}^\mathrm{opt}$ that leads to the best approximation of $\vec{H}_\mathrm{red}$ by using \eqref{eq:Hred2}. This is identical to the following optimization problem that determines the optimal susceptances $\vec{b}^\mathrm{opt}$ in the reduced network by minimizing the Frobenius matrix norm:
\begin{equation}
\begin{array}{l}
\displaystyle\min_{\vec{b}^\mathrm{opt}} \ \left \| {\vec{H}_\mathrm{red}}  - \mathrm{diag}\{\vec{b}^\mathrm{opt}\}\vec{C} \left(\vec{C}^T\mathrm{diag}\{\vec{b}^\mathrm{opt}\}\vec{C} \right)^{-1} \right\|_{\mathrm{F}}^2\\
\text{s.t.} \\
\text{(a)} \ \vec{a}^T \vec{b}^\mathrm{opt} = b_m \quad.
\label{eq:redOpt} 
\end{array}
\end{equation}
Note that problem~\eqref{eq:redOpt} is nonlinear, since $\vec{b}^\mathrm{opt}$ is part of an inverse matrix. The difference to the aforementioned methods is that we enforce a physical meaningful solution by incorporating $\vec{a}$ and $b_m$ from Section~\ref{sec:lsfit} as a hard constraint with (\ref{eq:redOpt}a). This is advantageous because this approach does not influence the objective and therefore avoid a suboptimal solution that could be retrieved by the method from Section~\ref{sec:lsfit}.

\section{Results}
\label{sec:results}
This section presents a comparison of the introduced network reduction methods. We compare the power flow errors that occur in both calculation stages (see Fig~\ref{fig:process}). This includes the reduced PTDF matrices from the first stage and the B approximation methods from the second stage. The results are structured into two parts. For the sake of comprehensiveness the first part gives illustratively detailed results on the reduced PTDF matrices and susceptances for the IEEE14-bus system. The second part is primarily concerned with comparing the power flow errors and computation time for the considered methods in dependence of differently-sized grids. 
\subsection{Error Metric}
For comparison reasons, we need to define an error metric that can be used for the case studies. We specify the Normalized Root Mean Square Error (NRMSE) as 
\begin{equation}
\text{NRMSE} = \frac{\text{RMSE}\left(\vec{T}_\mathrm{f}\vec{H}_\mathrm{f}\vec{p}_\mathrm{inj} - \vec{H}_y^x\vec{T}_\mathrm{bz}\vec{p}_\mathrm{inj}\right)} {\text{mean}\left(|\vec{T}_\mathrm{f}\vec{H}_\mathrm{f}\vec{p}_\mathrm{inj}|\right)} \quad, \label{eq:error}
\end{equation}

\noindent which can be regarded as the average relative power flow deviation per line between the original aggregated flows and the reduced flows. The function `RMSE' is the root mean square error.

\subsection{Implementation}
All methods are implemented in MATLAB on a Computer with an Intel Core i7-6600U processor. The function \texttt{fmincon} solves the nonlinear problem \eqref{eq:redOpt} at the initial point $\vec{b}^\mathrm{phys}$. We compute the power flow errors for the different approaches according to \eqref{eq:error}. We run this error analysis over a given set of different power injection scenarios ($n$ = 3000) from which we obtain a power flow NRMSE distribution. From this distribution we calculate the mean value $\overline{\text{NRMSE}}$. The power injection samples are drawn from a normal distribution.

\subsection{IEEE14-Bus System}
\label{sec:case14}
\begin{figure}[t]
	\centering
	\includegraphics[width = \columnwidth]{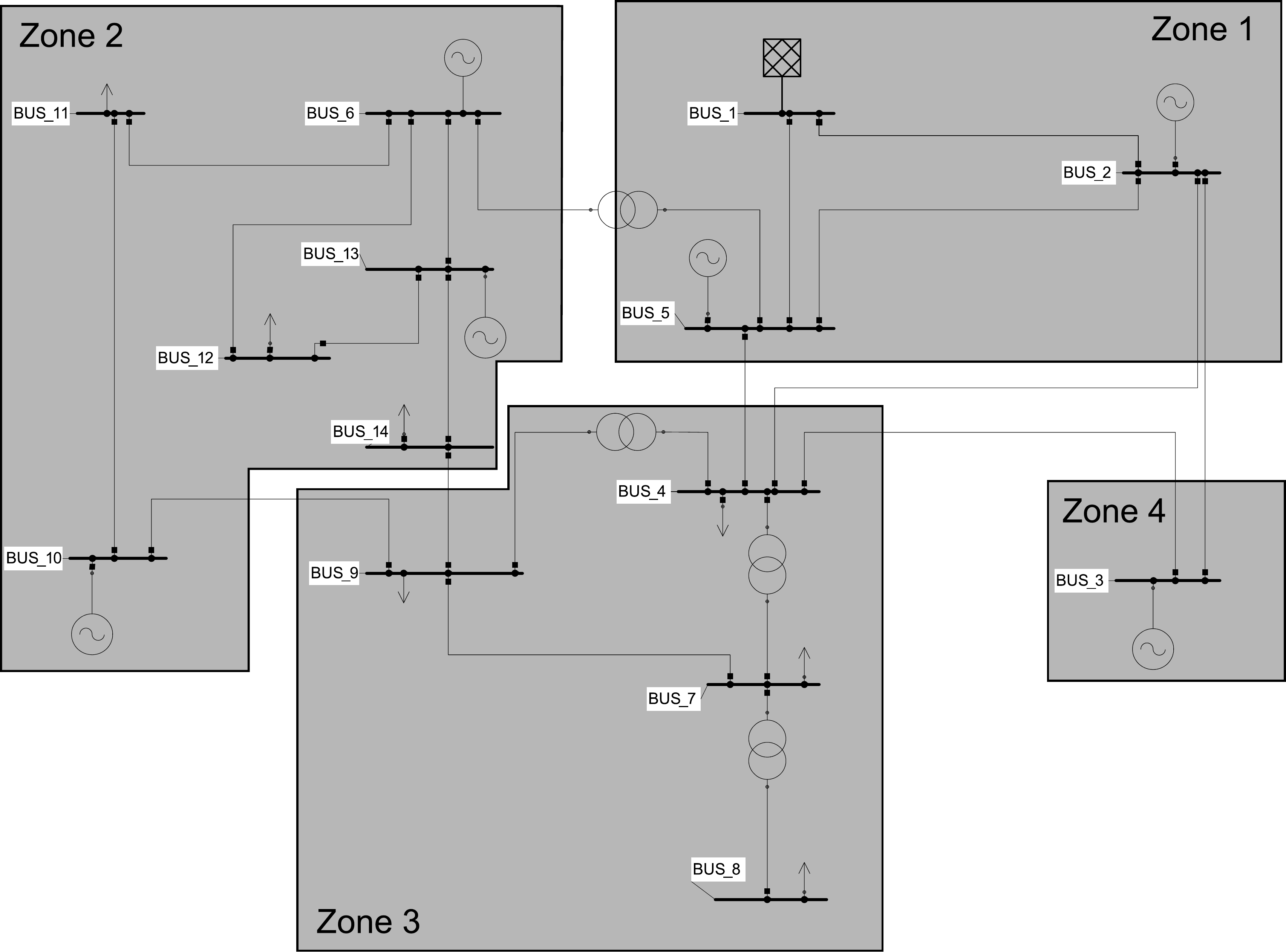}
	\caption{Division of the IEEE14-bus system into four zones for the first case study.}
	\label{fig:ieee14_zones}
\end{figure}

As depicted in Fig.~\ref{fig:ieee14_zones}, we divide the IEEE14 grid into 4 zones. The nodal injections are set according to Table~\ref{tab:zones}. 

\begin{table}
	\caption{Nodal injection configuration and zonal assignment for the IEEE14-bus system.}
	\centering
	\begin{tabular}{rrr}
		\toprule
		Bus & \multicolumn{1}{c}{ $p_{\mathrm{inj}}$ (MW) }& Zone \\
		\midrule
		1 (slackbus) & 41 & 1 (slackbus) \\
		2 & 46 & 1\\
		3 & 37 & 4\\
		4 & -57 & 3\\
		5 & 34 & 1\\
		6 & 13 & 2\\
		7 & -94 & 3\\
		8 & -20 & 3\\
		9 & -22 & 3\\
		10& 61 & 2\\
		11 & -27 & 2\\
		12 & -21 & 2\\
		13 & 13 & 2\\
		14 & -4	 & 2\\	
		\bottomrule
	\end{tabular}
	\label{tab:zones}
\end{table}

From the test grid in Fig.~\ref{fig:ieee14_zones} we construct the following mapping matrices in the index space of the original branches labeled column-wise and the interzonal connections labeled row-wise for 
\begin{equation}
\vec{T}_\mathrm{f}=
\begin{blockarray}{l@{\hspace{0pt}} *{20}{c@{\hspace{2pt}}} }
	\begin{block}{l@{\hspace{0pt}} *{20}{>{$\tiny}r<{$}@{\hspace{2pt}}} }
		 & 1-2 & 1-5 & 2-3 & 2-4 & 2-5 & 3-4 & 4-5 & 4-7 & 4-9 & 5-6 & 6-11 & 6-12 & 6-13 & 7-8 & 7-9 & 9-10& 9-14 & 10-11&12-13 & 13-14\\
	\end{block}
	\begin{block}{>{$\tiny}l@{\hspace{6pt}}<{$} [*{20}{c@{\hspace{2pt}}}]}
	1-4 &  \bigstrut[t] 0 & 0 & 1 & 0 & 0 & 0 & 0 & 0 & 0 &  0& 0 & 0 & 0 & 0& 0& 0 & 0  & 0 & 0 & 0  \\
	1-3 &			   0 & 0 & 0 & 1 & 0 & 0 & \text{-}1 & 0 & 0&0&0&0&0&0& 0&0&0&0&0&0\\
	4-3 &			   0 & 0 & 0 & 0 & 0 & 1 & 0 &0 &0 &0 &0 &0 &0&0&0&0&0&0&0&0 \\
	1-2 & 			   0 & 0 & 0 &0 &0 &0 & 0 & 0 & 0 &1 & 0 &0&0&0&0 &0 &0 &0 &0 &0\\
	3-2 & 			   0 &0&0&0&0&0&0&0&0&0&0&0&0&0&0&1&1&0&0&0\\
	\end{block}
\end{blockarray} 
\end{equation}
\noindent and in the index space of the original buses labeled column-wise and the zones labeled row-wise for 
\begin{equation}
\vec{T}_\mathrm{bz} =
\begin{blockarray}{l@{\hspace{0pt}} *{13}{c@{\hspace{2pt}}} }
\begin{block}{l@{\hspace{0pt}} *{13}{>{$\tiny}r<{$}@{\hspace{2pt}}} }
&  2 & 3 & 4 & 5& 6 & 7 & 8 & 9 &10& 11 & 12 & 13 & 14\\
\end{block}
\begin{block}{>{$\tiny}l@{\hspace{5pt}}<{$} [*{13}{c@{\hspace{2pt}}}]}
2  & \bigstrut[t] 0 & 0& 0 & 0 & 1 &0 &0&0 & 1&1&1&1&1 \\
3 &			  0 & 0& 1 & 0 &0 & 1&1&1&0&0&0&0&0\\
4 &			  0 & 1 & 0 &0&0&0&0&0&0&0&0&0&0 \\
\end{block}
\end{blockarray} \quad.
\end{equation}
\noindent

\subsubsection{Reduced PTDF Comparison}
Now, with the aforementioned defined matrices, we calculate the reduced PTDF matrices according to \eqref{eq:redPTDFshi} and \eqref{eq:redPTDFoh} for the injection-dependent and the injection-independent approaches that are determined in the first stage from Fig.~\ref{fig:process}. The results are listed in Table~\ref{tab:redPTDF}.
\begin{table}[t]
	\caption{Reduced PTDF results for the IEEE14-bus system.}
	\centering
	\begin{tabular}{p{1.5cm}p{5.5cm}} 
		\toprule
		Approach &  \multicolumn{1}{c}{PTDF} \\
		\midrule
		injection-dependent \cite{Shi2015} & $\vec{H}_\mathrm{red}^\mathrm{dep} = \left[ \begin{array}{rrr}  -0.138 & -0.144 & -0.532 \\
																							   -0.582 & -0.695 & -0.450 \\
																							  -0.138 & -0.144 & 0.468 \\
																							   -0.278 & -0.159 & -0.017 \\
																							   -0.721 & 0.159 &  0.017\end{array}\right] $ \\
		\hline
		injection-independent \cite{Oh2010} & $\vec{H}_\mathrm{red}^\mathrm{ind} = \left[ \begin{array}{rrr}  -0.126 & -0.143 & -0.532 \\
																							    -0.343 & -0.676 & -0.450  \\
																							     -0.126 & -0.143 & 0.468 \\
																							     -0.530 & -0.179 & -0.017 \\
																							    -0.469 & 0.179 & 0.017 \end{array}\right]$ \\
		\bottomrule
	\end{tabular}
	\label{tab:redPTDF}
\end{table}
\begin{figure}[t]
	\centering	
	\subfloat[Zonal power flows for the specified fixed nodal bus injections.]{
	\def\svgwidth{\columnwidth}
	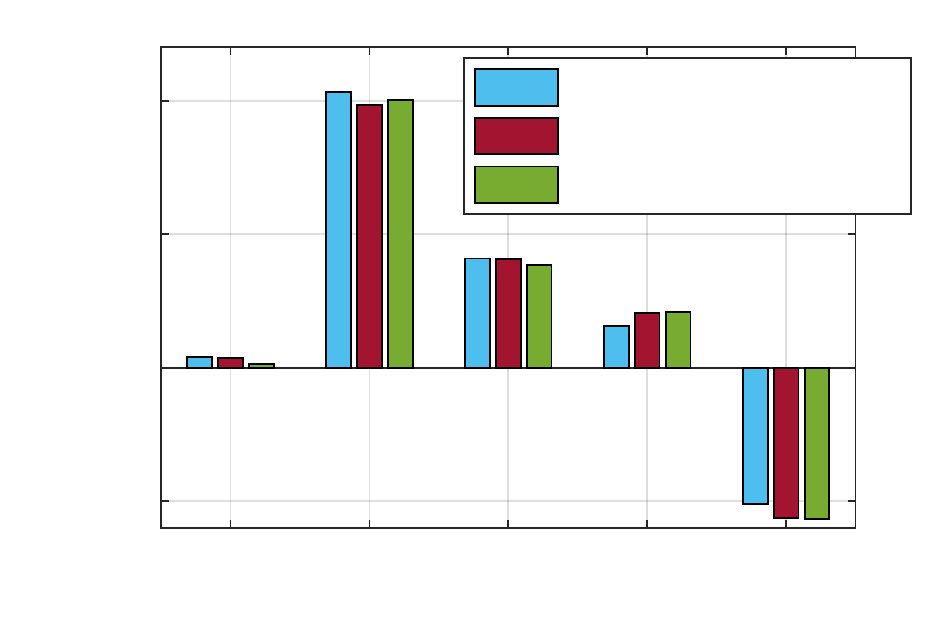
	\label{fig:ptdf_flow_compare}} \\
	\subfloat[Power flow error comparison using 3000 arbitrary selected power injection scenarios. The bars on the NRMSEs indicate the mean values of the associated NRMSE distribution.]{
	\def\svgwidth{\columnwidth}
	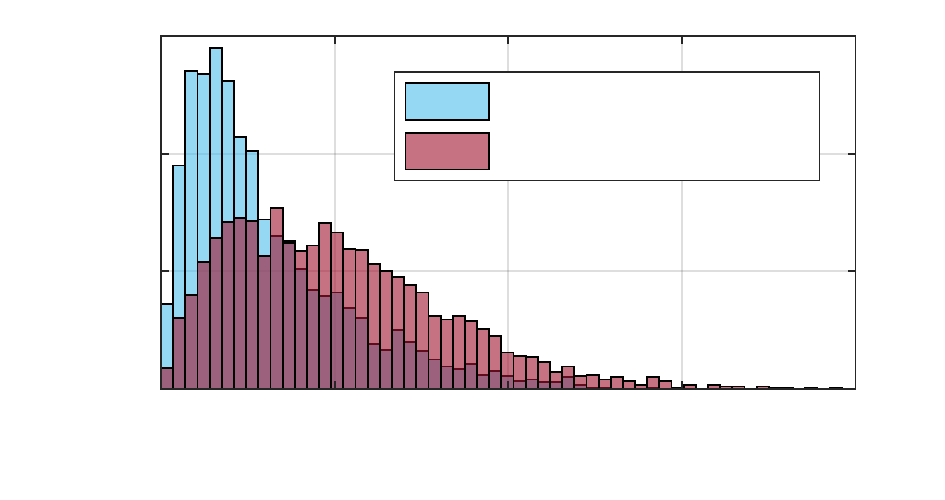 \label{fig:ptdf_compare}}
	\caption{Reduced PTDF comparison for the IEEE14-bus system.}
	\vspace{-0.2cm}
\end{figure}

Next, we compare the interzonal power flows by applying \eqref{eq:p_red2} between different reduced PTDF approximations for the fixed power injection listed in Table~\ref{tab:zones}. We also calculate the power flow errors using \eqref{eq:error} for the different approximations. Figure~\ref{fig:ptdf_flow_compare} shows the results. From this it can be observed that the injection-dependent PTDF (`dark red bars') achieves the best approximation indicating with the lowest power flow error (3.8\%), while the injection-independent (`blue bars') perform worse. 

If we run the same analysis over a set of different power injection scenarios ($n$ = 3000), we obtain a power flow NRMSE distribution as shown in Fig.~\ref{fig:ptdf_compare}. As a result by comparing the $\overline{\text{NRMSE}}$, we find  that in this case the injection-independent version performs here by nearly a factor of 2 better than the injection-dependent version. This can be explained that the calculation of the injection-independent PTDF minimizes the average over all possible injections resulting in a lower error, while at the same time having potentially higher errors for a fixed power injection scenario.

\subsubsection{Susceptance Comparison}
In the second stage, we calculate the susceptances and perform a power flow error analysis for the susceptance (B) approximation methods presented in Section~\ref{sec:detSus}. As specified in \cite{Shi2015} we consider the injection-dependent PTDF matrix for the $\vec{b}^\mathrm{shi}$ determination, while we use the injection-independent version for the nonlinear optimization-based $\vec{b}^\mathrm{opt}$ and $\vec{b}^\mathrm{oh}$ determination. The results are given in Table~\ref{tab:B}. The values of~$\vec{b}^{\mathrm{oh}}$ differ significantly from the other methods. This can be explained that this method does not enforce a physical meaningful solution. However, since any constant applied to the solution does not affect the reduced PTDF matrix, a meaningful solution can be retrieved a posteriori. Note that it might occur that some susceptances could take negative values that would correspond to capacitive line reactances. For uniform solutions we adjust capacitive to inductive line reactances by changing them to positive values. Apart from the voltage angle differences the DC power flow solution remains unchanged under this adjustment in this case.  
\begin{table}[t]
	\caption{Susceptance results for the presented approximation methods.}
	\centering
	\begin{tabular}{rrrrr} 
		\toprule
		& \multicolumn{4}{c}{B Approximation method} \\
		& \multicolumn{1}{c}{physical } & \multicolumn{1}{c}{Shi \cite{Shi2015} } & \multicolumn{1}{c}{Oh \cite{Oh2010} }  & \multicolumn{1}{c}{optimal } \\
		& $\vec{b}^\mathrm{phys}$ & $\vec{b}^\mathrm{shi}$ & $\vec{b}^\mathrm{oh}$ & $\vec{b}^\mathrm{opt}$ \\
		\midrule
		$b_{14}$ & 5.05 & 3.53 &  0.19 & 12.47\\
		$b_{13}$ & 29.41 & 9.26 & 0.71& 29.41\\
		$b_{43}$ & 5.84 & 2.96 & 0.25& 16.97\\
		$b_{12}$ & 3.96 & 4.00 & 0.37& 11.04\\
		$b_{32}$ & 15.53 & 2.81 & 0.48& 12.98\\
		\bottomrule
	\end{tabular}
	\label{tab:B}
\end{table}
\begin{figure}[t]
	\centering
	\subfloat[Zonal power flows for the specified fixed nodal bus injections.]{%
		\def\svgwidth{\columnwidth}
		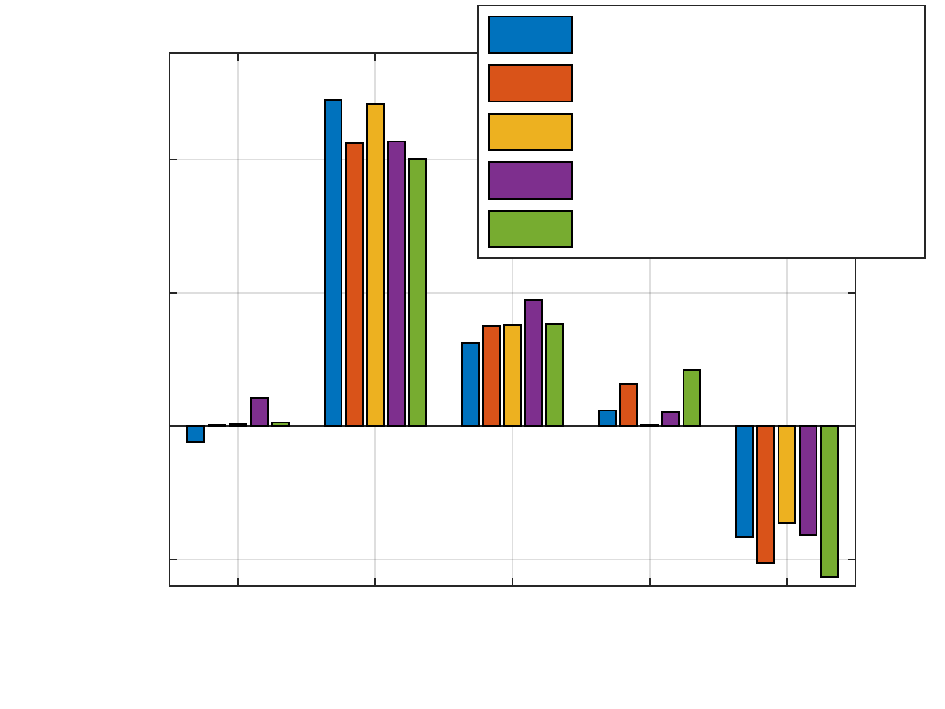 \label{fig:B_flow_compare}} \\
	\subfloat[Power flow error comparison using 3000 arbitrary selected scenarios. The bars on the NRMSEs indicate the mean values of the associated NRMSE distribution.]{%
		\def\svgwidth{\columnwidth}
		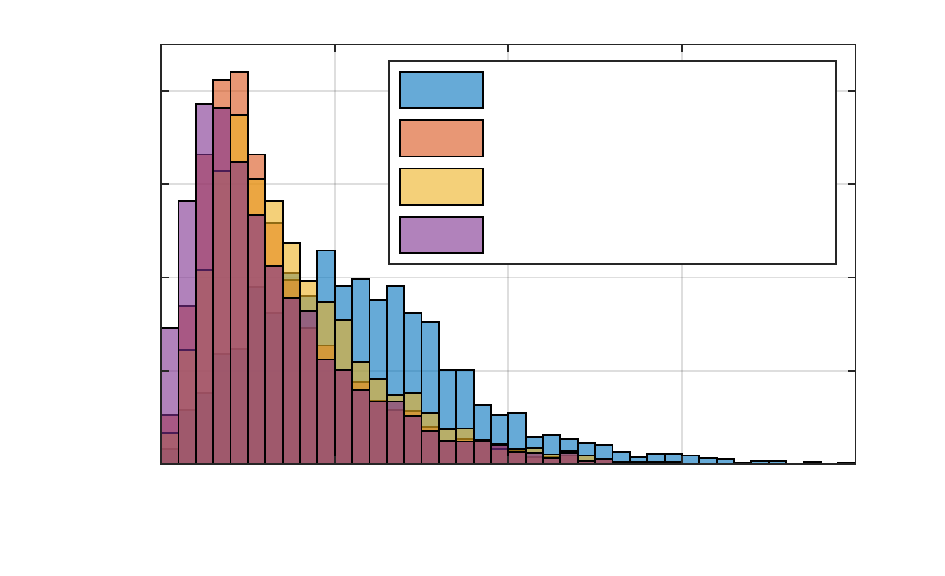 \label{fig:B_compare}} \\
	\caption{B approximation method comparison for the IEEE14-bus system.}
	\vspace{-0.2cm}
\end{figure}
\begin{table*}[t]
	\caption{Results for different grid cases.}
	\centering
	\setlength{\tabcolsep}{12pt}
	\begin{tabular}{@{}>{\rm\PBS\raggedright}p{1cm}@{} %
			@{}>{\rm\PBS\raggedleft}p{0.7cm}@{}
			@{}>{\rm\PBS\raggedleft}p{0.7cm}@{}
			@{}>{\rm\PBS\raggedleft}p{0.7cm}@{}
			@{}>{\rm\PBS\raggedleft}p{0.7cm}@{}
			@{}>{\rm\PBS\raggedright}p{1cm}@{}
			@{}>{\rm\PBS\raggedright}p{1cm}@{}
			@{}>{\rm\PBS\raggedright}p{1cm}@{}
			@{}>{\rm\PBS\raggedright}p{1cm}@{}
			@{}>{\rm\PBS\raggedright}p{1cm}@{}
			@{}>{\rm\PBS\raggedright}p{1cm}@{}
			@{}>{\rm\PBS\raggedright}p{1cm}@{}
			@{}>{\rm\PBS\raggedright}p{1cm}@{}
			@{}>{\rm\PBS\raggedright}p{1cm}@{}}
		\toprule
		Grid &  \multicolumn{2}{c}{Buses} & \multicolumn{2}{c}{Branches} &  \multicolumn{6}{c}{$\overline{\text{NRMSE}}$} &  \multicolumn{3}{c}{Calculation Time (sec)}  \\
		\cmidrule(l){2-3} 
		\cmidrule(lr){4-5} 
		\cmidrule(lr){6-11} 
		\cmidrule(lr){12-14} 
		& 	$n_b$	& $n_z$	&	$n_l$ & $n_l^r$	& $\vec{H}_\mathrm{red}^\mathrm{ind}$ &  $\vec{H}_\mathrm{red}^\mathrm{dep}$ & $\vec{H}_\mathrm{b}^\mathrm{phys}$ & $\vec{H}_\mathrm{b}^\mathrm{oh}$ &  $\vec{H}_\mathrm{b}^\mathrm{shi}$ &  $\vec{H}_\mathrm{b}^\mathrm{opt}$ &  $\vec{H}_\mathrm{b}^\mathrm{oh}$ &  $\vec{H}_\mathrm{b}^\mathrm{shi}$ &  $\vec{H}_\mathrm{b}^\mathrm{opt}$ \\
		\midrule
		6bus & 6 & 4 & 6 & 5 & 0.24 & 0.25 & 0.26 &0.24 & 0.32 & 0.24 & 0.021 & 0.0043 & 0.24 \\
		ieee14 & 14 & 4 & 20 & 5 & 0.30 & 0.51 & 0.57 & 0.33 & 0.38 & 0.31 & 0.0059 & 0.0026 & 0.22 \\
		ieee39 & 39 & 3 & 46 & 3 & 0.25 & 0.38 & 0.26 & 0.25 & 0.26 & 0.25 & 0.011 & 0.00023 & 0.15 \\
		185bus & 185 & 21 & 352 & 41 &  0.51 & 1.05 & 0.89 & 0.72 & 1.15 & 0.56 & 0.0063 & 0.0014 & 5.06 \\
		\midrule
		\multicolumn{5}{r}{Average} &  0.32 & 0.54 &  0.49 & 0.38 & 0.52 & 0.34 \\
		\cmidrule{6-10}
		\multicolumn{5}{r}{Average improvement of $\vec{H}_\mathrm{b}^\mathrm{opt}$} & -6\% & +58\% & +44\% & +11\% & +52\%  \\
		\cmidrule{7-11}
		\multicolumn{5}{r}{Average improvement of $\vec{H}_\mathrm{red}^\mathrm{ind}$} &   & +68\% & +53\% & +18\% & +62\% & +6\%  \\
		\midrule
		wp2746 & 2746 & 6 & 3514& 11 & 0.69 & 2.23 & 2.01 & 1.5 & 1.88 & 1.43 & 0.001 & 0.0044 & 0.128 \\
		\bottomrule
	\end{tabular}
	\label{tab:griderrors}
\end{table*}
If we run a power flow error analysis for the fixed power injection (Fig.~\ref{fig:B_flow_compare}), we find that the B approximation from~\cite{Shi2015} incurs the highest error (36\%) and the B approximation from~\cite{Oh2010} achieves the lowest error (9.9\%). However, this picture changes when we look over a given set of power injection scenarios (Fig.~\ref{fig:B_compare}). The physical-based B approximation method is the worst (57\%), while the B approximation method based on nonlinear optimization (`purple bars') achieves the best result with 31\% power flow error. This is only 1\% point more than compared to the injection-independent PTDF. In general, the errors of the B approximation methods are higher than for the injection-independent PTDF method, since an additional error is generated by the second stage of the network reduction process~(see Fig.~\ref{fig:process}).

\subsection{Different Grid Cases}
The findings of the previous Section~\ref{sec:case14} refer to one grid, which might not be representative. To generalize our statements, we consolidate our findings for a larger set of differently-sized grids. We use grids that are bundled in the MATPOWER framework \cite{matpower} and are already divided into zones. The power flow errors are also dependent on how the zones are composed. Optimal divisions can be obtained by using k-means clustering \cite{Shi2015}.  However, we do not consider an optimal clustering here, since for several applications the network division can be either geographically or politically fixed. In Table~\ref{tab:griderrors} the different grids are characterized by their name, buses, and branches. The averaged power flow errors are calculated based on a set of scenario ($n$=3000) simulations and are also listed for the different suggested approaches and grids. 

\subsubsection{Power Flow Approximation Errors}
\begin{figure}[t]
	\centering
	\def\svgwidth{\columnwidth}
	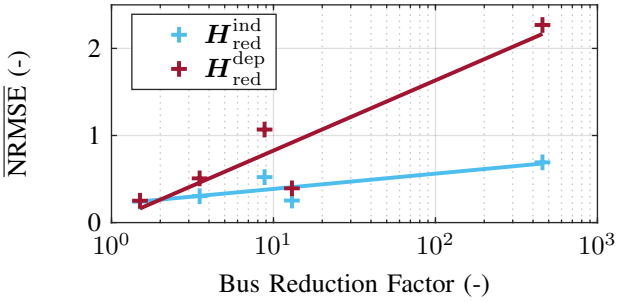
	\caption{Power flow error as a function of the bus reduction for the suggested reduced PTDF matrices. The lines represent the associated linear regressions.}
	\label{fig:ptdf_compare_grids}
\end{figure}
\begin{figure}[t]
	\centering
	\def\svgwidth{\columnwidth}
	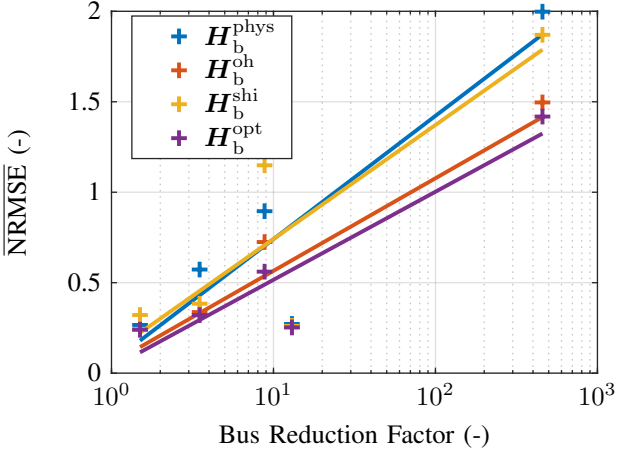
	\caption{Power flow error as a function of the bus reduction. The lines correspond to the linear regressions associated with the different B approximation methods.}
	\label{fig:B_compare_grids}
\end{figure}

From the average power flow errors in Table~\ref{tab:griderrors} we can conclude that taking the injection-independent PTDF is in average better than using the injection-dependent PTDF (error improvement by 68\%). Overall, one can say that the best approximation is obtained by the injection-independent PTDF approximation (mean error 32\%). Moreover, our suggested nonlinear optimal B approximation method outperforms all other B approximation methods by 11\% at minimum. Note that we take out the wp2746 results in the average calculation, since the bus reduction ratio is too high to represent a reasonable compact system. Also we can observe that the additional error that is imposed by the second stage in the network reduction process amounts to at least 6\% if we apply the nonlinear optimal B approximation method.  

From the results in Table~\ref{tab:griderrors} we could also identify the power flow error as a function of the bus reduction ratio. This is important when we want to find a trade-off between the dimension of the reduced network and the powerflow error. We rearrange our results as a function of the bus reduction ratio that is defined as  
\begin{equation}
\text{Bus Reduction Ratio} = \frac{n_b}{n_z} \quad,
\end{equation}
\noindent and show them in the Figures~\ref{fig:ptdf_compare_grids} and \ref{fig:B_compare_grids}. The high fluctuations of the NRMSEs indicate that the zonal division has a great impact on the power flow errors. This means that these results should be interpreted as an upper bound for the power flow errors. The relative power flow error at bus reduction ratios above 100 already constitutes 100\%. This means that a faithful conversion cannot be given above this ratio.     

\subsubsection{Computation Time}
\begin{figure}[t]
	\centering
	\def\svgwidth{\columnwidth}
	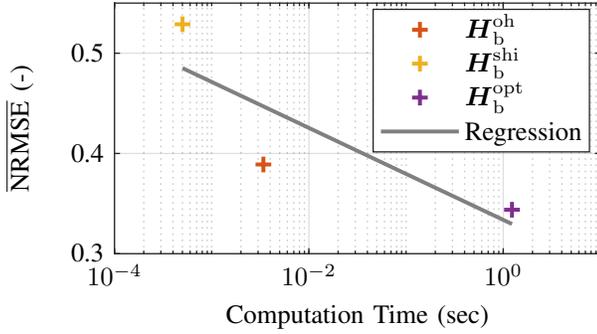
	\caption{Trade-off between the averaged power flow error and the computation time that is averaged for the different B approximation methods. The gray line indicates the lin-log regression.}
	\label{fig:comp_time_grids}
\end{figure}

Lastly, we provide a comparison on the computation times for the different B approximation approaches that take a reduced PTDF as input. It should be mentioned that the calculation time for the first stage is much longer than for the second one, since the time for evaluating the full PTDF $\vec{H}_\mathrm{f}$ with \eqref{eq:Hf1} or \eqref{eq:Hf2} is the dominating factor. However, there are still time differences in the second stage that are present for the different B approximation approaches. Its computation time depends on the number of the interzonal connections, because the dimensions of the matrices to determine the susceptances as presented in Section~\ref{sec:detSus} contain the number of the zonal interconnections. Figure~\ref{fig:comp_time_grids} shows the trade-off between the averaged computation time of the presented B approximation methods and their averaged power flow errors. It can be observed that the computation time of our nonlinear optimization based B approximation method takes longer than for the other presented methods, but at the same time it achieves the lowest power flow error.

\section{Conclusion}
\label{sec:conclusion}
In this paper several methods are presented to obtain a reduced representation of a power transmission network. One way to convert the original system into a smaller network is to project the reduced PTDF matrix on the original one. We present two methods from literature to obtain the reduced PTDF matrix. Although an injection-dependent PTDF matrix achieves lower approximation errors for a fixed power injection scenario, we could find that under a given set of arbitrarily defined power injection scenarios an injection-dependent reduced PTDF matrix introduces on average higher power flow errors as compared to an injection-independent version. We found an upper bound for the bus reduction ratio, at which a faithful network conversion cannot be given any longer.

To get a sparse representation of the network the susceptances of the reduced network have to be determined. This second stage imposes an additional power flow error. We propose a nonlinear optimization-based susceptance approximation method and compare our developed method with two methods from literature. We showed that our method reduces the power flow error as compared to those from literature and has a comparable power flow error as the injection-independent reduced PTDF matrix from the first stage. It can be anticipated that the higher computation time of our method is reasonable and not critical for the considered applications.

\section*{Acknowledgement}
This research is part of the activities of Nexus -- an integrated energy systems modeling platform, which is financially supported by the Swiss Federal Office of Energy (SFOE).

\bibliographystyle{IEEEtran}
\bibliography{literature}

\begin{thebibliography}{10}
\providecommand{\url}[1]{#1}
\csname url@samestyle\endcsname
\providecommand{\newblock}{\relax}
\providecommand{\bibinfo}[2]{#2}
\providecommand{\BIBentrySTDinterwordspacing}{\spaceskip=0pt\relax}
\providecommand{\BIBentryALTinterwordstretchfactor}{4}
\providecommand{\BIBentryALTinterwordspacing}{\spaceskip=\fontdimen2\font plus
\BIBentryALTinterwordstretchfactor\fontdimen3\font minus
  \fontdimen4\font\relax}
\providecommand{\BIBforeignlanguage}[2]{{%
\expandafter\ifx\csname l@#1\endcsname\relax
\typeout{** WARNING: IEEEtran.bst: No hyphenation pattern has been}%
\typeout{** loaded for the language `#1'. Using the pattern for}%
\typeout{** the default language instead.}%
\else
\language=\csname l@#1\endcsname
\fi
#2}}
\providecommand{\BIBdecl}{\relax}
\BIBdecl

\bibitem{DelaTorre2008}
S.~de~la Torre, A.~J. Conejo, and J.~Contreras, ``Transmission expansion
  planning in electricity markets,'' \emph{IEEE Transactions on Power Systems},
  vol.~23, no.~1, pp. 238--248, 2008.

\bibitem{Baringo2012}
L.~Baringo and A.~J. Conejo, ``Transmission and wind power investment,''
  \emph{IEEE Transactions on Power Systems}, vol.~27, no.~2, pp. 885--893,
  2012.

\bibitem{Roh2007}
J.~H. Roh, M.~Shahidehpour, and Y.~Fu, ``Market-based coordination of
  transmission and generation capacity planning,'' \emph{IEEE Transactions on
  Power Systems}, vol.~22, no.~4, pp. 1406--1419, 2007.

\bibitem{AlvarezLopez2007}
J.~{Alvarez Lopez}, K.~Ponnambalam, and V.~H. Quintana, ``Generation and
  transmission expansion under risk using stochastic programming,'' \emph{IEEE
  Transactions on Power Systems}, vol.~22, no.~3, pp. 1369--1378, 2007.

\bibitem{Murillo-Sannchez2013}
C.~E. Murillo-Sanchez, R.~D. Zimmerman, C.~{Lindsay Anderson}, and R.~J.
  Thomas, ``Secure planning and operations of systems with stochastic sources,
  energy storage, and active demand,'' \emph{IEEE Transactions on Smart Grid},
  vol.~4, no.~4, pp. 2220--2229, 2013.

\bibitem{Papavasiliou2013}
A.~Papavasiliou and S.~S. Oren, ``Multiarea stochastic unit commitment for high
  wind penetration in a transmission constrained network,'' \emph{Operations
  Research}, vol.~61, no.~3, pp. 578--592, 2013.

\bibitem{isgt}
P.~Fortenbacher, A.~Ulbig, S.~Koch, and G.~Andersson, ``Grid-constrained
  optimal predictive power dispatch in large multi-level power systems with
  renewable energy sources, and storage devices,'' in \emph{IEEE PES Innovative
  Smart Grid Technologies, Europe}, Oct 2014, pp. 1--6.

\bibitem{Fortenbacher2017}
P.~Fortenbacher, A.~Ulbig, and G.~Andersson, ``Optimal placement and sizing of
  distributed battery storage in low voltage grids using receding horizon
  control strategies,'' \emph{IEEE Transactions on Power Systems}, vol.~PP,
  no.~99, pp. 1--1, 2017.

\bibitem{Bergh2015}
K.~{Van den Bergh}, J.~Boury, and E.~Delarue, ``The flow-based market coupling
  in central western europe: Concepts and definitions,'' \emph{Electricity
  Journal}, vol.~29, no.~1, pp. 24--29, 2016.

\bibitem{Ward1949}
J.~B. Ward, ``Equivalent circuits for power-flow studies,'' \emph{Transactions
  of the American Institute of Electrical Engineers}, vol.~68, no.~1, pp.
  373--382, 1949.

\bibitem{Tinney1987}
W.~F. Tinney and J.~M. Bright, ``Adaptive reductions for power flow
  equivalents,'' \emph{IEEE Transactions on Power Systems}, vol.~2, no.~2, pp.
  351--359, May 1987.

\bibitem{Shi2012}
D.~Shi, D.~L. Shawhan, N.~Li, D.~J. Tylavsky, J.~T. Taber, R.~D. Zimmerman, and
  W.~D. Schulze, ``Optimal generation investment planning: Pt. 1: Network
  equivalents,'' \emph{2012 North American Power Symposium, NAPS 2012}, 2012.

\bibitem{Cheng2005}
X.~Cheng and T.~J. Overbye, ``{PTDF}-based power system equivalents,''
  \emph{IEEE Transactions on Power Systems}, vol.~20, no.~4, pp. 1868--1876,
  2005.

\bibitem{Oh2010}
H.~Oh, ``A new network reduction methodology for power system planning
  studies,'' \emph{{IEEE Transactions on Power Systems}}, vol.~25, no.~2, pp.
  677--684, 2010.

\bibitem{Shi2015}
D.~Shi and D.~J. Tylavsky, ``A novel bus-aggregation-based structure-
  preserving power system equivalent,'' \emph{IEEE Transactions on Power
  Systems}, vol.~30, no.~4, pp. 1977--1986, 2015.

\bibitem{matpower}
R.~D. Zimmerman, C.~E. Murillo-Sanchez, and R.~J. Thomas, ``Matpower:
  Steady-state operations, planning, and analysis tools for power systems
  research and education,'' \emph{IEEE Transactions on Power Systems}, vol.~26,
  no.~1, pp. 12--19, Feb 2011.

\end{thebibliography}

\end{document}